# Dynamic Power Distribution System Management With a Locally Connected Communication Network

Kaiqing Zhang, *Student Member, IEEE*, Wei Shi, *Member, IEEE*, Hao Zhu, *Member, IEEE*,
Emiliano Dall'Anese, *Member, IEEE*, and Tamer Başar, *Life Fellow, IEEE*

*Abstract*—Coordinated optimization and control of distribution-level assets can enable a reliable and optimal integration of massive amount of distributed energy resources (DERs) and facilitate distribution system management (DSM). Accordingly, the objective is to coordinate the power injection at the DERs to maintain certain quantities across the network, e.g., voltage magnitude, line flows, or line losses, to be close to a desired profile. By and large, the performance of the DSM algorithms has been challenged by two factors: i) the possibly non-strongly connected communication network over DERs that hinders the coordination; ii) the dynamics of the real system caused by the DERs with heterogeneous capabilities, time-varying operating conditions, and real-time measurement mismatches. In this paper, we investigate the modeling and algorithm design and analysis with the consideration of these two factors. In particular, a game-theoretic characterization is first proposed to account for a locally connected communication network over DERs, along with the analysis of the existence and uniqueness of the Nash equilibrium (NE) therein. To achieve the equilibrium in a distributed fashion, a *projected-gradient*-based asynchronous DSM algorithm is then advocated. The algorithm performance, including the convergence speed and the tracking error, is analytically guaranteed under the dynamic setting. Extensive numerical tests on both synthetic and realistic cases corroborate the analytical results derived.

## I. Introduction

Operation and management of power distribution systems can benefit in terms of reliability, optimality, and resilience from a systematic coordination of distributed energy resources (DERs). Specifically, the DERs can be coordinated to control certain quantities across the network, such as voltage magnitude or power line flows, by controlling both their real and reactive powers injected into the system in a real-time fashion. This is usually formulated as a network-wide optimal power flow (OPF) problem. We refer to [1]–[4] and the references therein for recent efforts based on this formulation. To solve the OPF problem, central/distributed optimization algorithms such as primal-dual [1] and ADMM [4] are advocated. A majority of these algorithms assume a *strongly-connected* communication network of high-quality to exist, such that the local information (either decision variables or measurements) can diffuse across all DERs in the network. However, communication and networking technologies in current distribution systems are still under-deployed and have limited capabilities. For example, several popular technologies including fiber optic cable, powerline carrier, and point-to-point microwave, more or less suffer from the limitations such as unacceptable delay, bandwidth constraints, and high deployment cost [5]–[7]. Therefore, the practical implementation of these optimization-based algorithms has been greatly challenged.

In fact, some recent work has accounted for these communication limitations in designing optimization-based algorithms for distribution system management (DSM). For example, in [8], a hybrid voltage control strategy is developed to be cognizant to the instantaneous availability of communication links. [9] designs distributed voltage control algorithms using quantized communication between neighboring buses to adapt to the bandwidth constraint of communications. On the other hand, several *local* control schemes that require no information exchange among DERs are developed to cope with such limited communications. For example, to regulate the voltage magnitude, DERs can perform *feedback* control using only local voltage measurements [10]–[12], even under asynchronous updates [13]. However, it has been proven that the local schemes make *myopic decisions* and lead to loss of optimality of the DSM performance [10], [11]. It is shown in [14] that there exists a trade-off between performance optimality and communication complexity for the voltage control problem, as argued in general networked control systems [15], [16].

In this paper, we propose a *semi-local* control scheme under a *locally connected communication network* over DERs to better characterize the performance-communication trade-off. In particular, we consider the scenario where buses with DERs are partitioned into several *communication areas*, where information can be exchanged only within each area. This scenario generalizes the cases either in local control schemes where each DER itself is a communication area, or in distributed schemes with a strongly-connected communication network where all DERs belong to a single area. The performance achieved under such a locally connected communication network fills in the gaps between the two special cases and characterizes the value of communication links. In addition, the scenario could be especially useful for the operation of networked micro-grids [17], where the physically connected micro-grids are usually owned and controlled by different entities or controllers. Because of possible competition or privacy concerns, very limited information exchange occurs among controllers. Under such locally connected communication networks, our recent work [18] developed the first equilibrium-learning algorithm for voltage control, with provable convergence guarantees. We note that our work is not the first one

This work is partially supported by the National Science Foundation under Award Number ECCS-1610732 and the Power Systems Engineering Research Center (PSERC) Project S-70. K. Zhang and T. Başar are with the Coordinated Science Laboratory, University of Illinois at Urbana-Champaign ({kzhang66, basar1}@illinois.edu). W. Shi is with Arizona State University (wilbur.shi@asu.edu). H. Zhu is with Department of Electrical and Computer Engineering at the University of Texas at Austin (haozhu@utexas.edu). E. Dall'Anese is with National Renewable Energy Laboratory (NREL) (emiliano.dallanese@nrel.gov).

that has considered the partition of distribution networks into several clusters/communities; see e.g., [19]–[21]. However, the network partition in this earlier work was performed not in accordance with the disconnected communication graph, but rather in order to decompose the overall DSM problem for better computational efficiency. Moreover, few analytical results have been provided for this line of work, except for extensive case studies. Whereas here, we aim to quantify the value of communication links from an analytical perspective, by developing algorithms that have provable convergence guarantees.

To this end, we introduce a game-theoretic characterization, where the areas managed by different controllers are modeled as players in a strategic game and only consider self-interest for lack of communication with each other. Equilibrium analysis is then provided under such a scenario, constituting the first aspect of the present work. Based on this, we propose an equilibrium-learning algorithm that uses real-time measurements as the feedback-based local control schemes.

Nonetheless, most existing *feedback* control schemes that rely on real-time measurements, see [8], [22], are challenged by the inevitable uncertainty in real distribution systems with DERs. Three main sources of uncertainty exist in the real system: i) the heterogeneous hardware capabilities of DERs, leading to possibly asynchronous operation of them, especially without coordination among communication areas; ii) the time-varying problem setting due to the fluctuation of uncontrollable load and operating limits of DERs at each bus; iii) limited precision of DERs sensing hardware and inaccuracy of the model, resulting in non-negligible measurements mismatches. It is thus imperative to investigate the performance of DSM algorithms under these settings for the *plug-and-play* functionality required for flexible DER integration [23].

The idea of accommodating real-time measurements into primal-dual-type methods for DSM goes back to [24], wherein a centralized controller was developed based on projected-gradient methods. Online algorithms were developed in [25], [26], [27], [28], and [29] to find solutions of AC OPF problems, with [27] establishing results in terms of tracking of solutions of a time-varying linearized AC OPF and [29] in terms of tracking of solutions of a time-varying relaxed AC OPF. Hybrid voltage regulation methods were presented in [8], [22]. Recently, a projected-gradient method on the (static) power flow manifold was proposed in [30] and an online incentive-based algorithm was proposed in [31], and results on terms of tracking of solutions of a time-varying non-convex problem were established. Finally, [23] focuses on the impact of asynchronous update and time-varying operating conditions on the convergence speed and tracking error of projected-gradient-based algorithms.

Relative to the works cited above, the present paper offers the following contributions:

• We propose a game-theoretic characterization for the dynamic DSM with a locally connected communication network and analyze the existence and uniqueness of its Nash equilibrium (NE); in contrast, [24], [29], [30] propose centralized algorithms, [26]–[28], [31] can accommodate only a broadcast (i.e., star) communication strategy, whereas [8], [23] assume a strongly-connected communication network;

• We develop an equilibrium-learning algorithm that relies on real-time measurements from the system and requires information exchange only within each communication area;

• In the spirit of [28], [31], [32], we analyze the dynamic performance of the proposed algorithm subject to time-varying operating conditions and measurements mismatches. Moreover, we also analyze the performance of the proposed scheme under asynchronous updates.

The remainder of the paper is organized as follows. Section II introduces the modeling of general DSM problems with local communications. In Section III, we first consider a static setting and provide a game-theoretic characterization of the problem along with its equilibrium analysis. In Section IV, we design a distributed learning algorithm that converges to the NE using real-time measurements and investigate its two variants with asynchronous updates. A practical setting with time-varying operating conditions and measurement mismatches is considered when analyzing the tracking performance of the asynchronous updates in Section V. Extensive numerical results are reported in Section VI to verify the theoretical results, followed by concluding remarks in Section VII.

## II. MODELING AND PROBLEM FORMULATION

In this section, we introduce the power flow model in distribution networks, followed by the formulation of the dynamic DSM problem with a locally connected communication network.

### A. Dynamic distribution system management

Consider a power distribution network represented by a graph $(\mathcal{N}, \mathcal{E})$. Let $\mathcal{N} := \{0, \cdots, N\}$ be the set of nodes (node 0 represents the feeder head and its voltage is taken as a reference) and $\mathcal{E} := \{(i,j), \forall i, j \in \mathcal{N}\}$ be the set of line segments. The set $\mathcal{N}_p := \mathcal{N}/\{0\}$ includes all the buses that have DERs with controllable real and reactive powers, which are denoted as $p_j$ and $q_j$ for bus-$j$, $\forall j \in \mathcal{N}_p$, respectively. Let $\mathbf{x}_j := (p_j, q_j)^T \in \mathbb{R}^2$, and let $\mathbf{z}_j \in \mathbb{R}^m$ be a vector collecting pertinent measurable electrical quantities at bus-$j$. Vector $\mathbf{z}_j$ collects, for example, electrical states that can be measured utilizing existing meters or estimated using state estimation techniques; these states include voltage magnitudes, line power flows, and line currents. As we will discuss in Section V, the proposed algorithm will account for *measurements and estimation errors*. All local variables $\{\mathbf{x}_j\}$ and $\{\mathbf{z}_j\}$ are concatenated into vectors $\mathbf{x} \in \mathbb{R}^{2N}$ and $\mathbf{z} \in \mathbb{R}^{mN}$, respectively.

To facilitate the design of the dynamic algorithm along with its performance analysis, we leverage approximate power-flow models that linearly relate the state $\mathbf{z}$ with the DER powers $\mathbf{x}$. In particular, an approximate relationship can be represented as

$$\mathbf{z} = \mathbf{H}\mathbf{x} + \bar{\mathbf{z}} \quad (1)$$

where the matrix $\mathbf{H} \in \mathbb{R}^{mN \times 2N}$ and the vector $\bar{\mathbf{z}} \in \mathbb{R}^{mN}$ can be obtained as shown in e.g., [33] (see also pertinent references therein). Even though a linearized method is utilized for the synthesis of the algorithm, its performance will be assessed by accounting for the non-linear AC power flow equations [31]. Notice that matrix $\mathbf{H} \in \mathbb{R}^{mN \times 2N}$ captures the sensitivity of $\mathbf{z}$ with respect to $\mathbf{x}$ [34].

**Remark 1.** For exposition and notational simplicity, modeling and analysis in the rest of paper focus on single-phase distribution networks with one DER per node. However, the proposed approach can be straightforwardly extended to unbalanced multi-phase distribution systems [33] with multiple DERs per node.

**Remark 2.** It is worth emphasizing the following three aspects that justify the use of the linear model (1) here. First, the linear model facilitates the development of computationally-efficient algorithms that: (i) afford a real-time implementation; (ii) can be implemented using the communication infrastructure illustrated in Fig. 1; and, (iii) enjoy analytical convergence guarantees, as we will show shortly in the paper. Second, although a linear model is utilized, appropriate measurements (i.e., feedback) are utilized in the algorithmic steps to fully account for the *nonlinear* physics of the distribution network. In fact, approximation errors associated with the linear model (1) are accounted for in our convergence analysis as in Assumption 6. Third, several recent results have shown that the accuracy of linear models is quite competitive even for distribution systems; see e.g., [33], [34]. All these make a case for the linear modeling approach used in the present work.

Consider a dynamic DSM setting where the set-point needs to be updated at each time slot $t$, where $t \in \mathbb{N}$ is the time index with certain sampling period. At time $t$, the general DSM objective is to minimize the difference between the controlled quantity $\mathbf{z}^{(t)}$ and a desired profile $\boldsymbol{\mu}$, i.e., $\mathbf{z}^{(t)} \to \boldsymbol{\mu}$, while considering the cost of providing powers associated with a predefined function $C_j^{(t)} : \mathbb{R}^2 \to \mathbb{R}$. In particular, $C_j^{(t)}(p_j, q_j) := C_{p,j}^{(t)}(p_j) + C_{q,j}^{(t)}(q_j)$ where $C_{p,j}^{(t)}, C_{q,j}^{(t)} : \mathbb{R} \to \mathbb{R}$ are the time-varying cost functions for real and reactive power provision at bus-$j$, respectively. The operating point captured by $\bar{\mathbf{z}}^{(t)}$ can vary over time due to the inevitable dynamics of uncontrollable load and renewable generation. With these definitions in mind, we formulate the following *time-varying* optimization problem, denoted as $\mathcal{P}^{(t)}$, at each time slot $t$:

$$\mathcal{P}^{(t)}: \quad \min_{\mathbf{z},\mathbf{x}} \quad \frac{\eta}{2}\|\mathbf{z}-\boldsymbol{\mu}\|_2^2 + \gamma \sum_{j=1}^{N} C_j^{(t)}(\mathbf{x}_j) \quad (2a)$$

$$s.t. \quad \mathbf{z} = \mathbf{H}\mathbf{x} + \bar{\mathbf{z}}^{(t)} \quad (2b)$$

$$\mathbf{x}_j \in \mathcal{X}_j^{(t)}, \forall j \in \mathcal{N}_p \quad (2c)$$

where $\mathcal{X}_j^{(t)}$ denotes the convex and compact set of possible powers that can be provided at bus-$j$ based on the local DERs' physical capability. The set $\mathcal{X}_j$ usually has the form of either a box constraint with upper and lower limits $\bar{\mathbf{x}}_j$ and $\underline{\mathbf{x}}_j$, i.e., $\{\mathbf{x}_j | \underline{\mathbf{x}}_j \leq \mathbf{x}_j \leq \bar{\mathbf{x}}_j\}$, or a more complicated one with limited apparent power capacity $s_j^{max}$, i.e., $\{\mathbf{x}_j | \|\mathbf{x}_j\|_2^2 \leq s_j^{max}\}$. The coefficients $\eta > 0$ and $\gamma \geq 0$ are the parameters that balance the control mismatch and the power provision cost in the objective. The larger $\eta$ is compared with $\gamma$, the greater emphasis is laid on the regulation of $\mathbf{z}$ to $\boldsymbol{\mu}$ as opposed to the power provision cost. Here we make a standard assumption pertaining to the general cost functions.

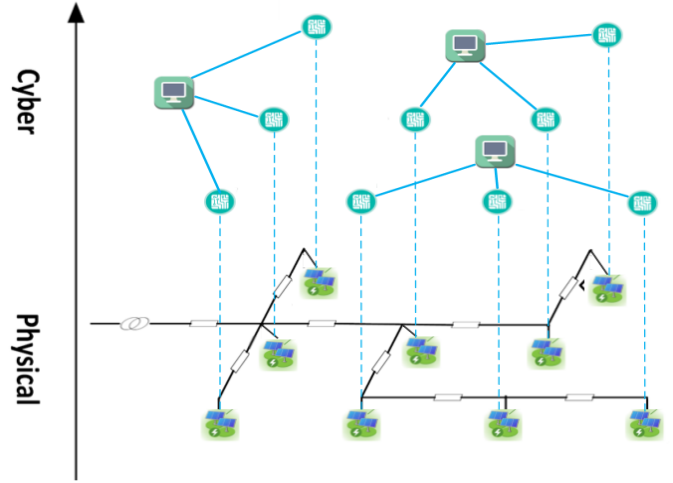

Fig. 1. A diagram for current distribution systems with DERs under a locally connected communication network. On the cyber layer, the blue lines represent the communication links connecting DER sensors and the central controller; on the physical layer, DERs are connected via distribution lines.

**Assumption 1.** *The cost functions $C_{p,j}^{(t)}$ and $C_{q,j}^{(t)}$ are $\nu$-strongly convex and have gradients that are L-Lipschitz continuous over the region $\mathbf{x}_j \in \mathcal{X}_j$ for each $j \in \mathcal{N}_p$ and $t \in \mathbb{N}$.*

### B. Locally connected communication network

As shown in Fig. 1, we consider a communication network $(\mathcal{N}, \mathcal{E}_c)$ deployed for the distribution network $(\mathcal{N}, \mathcal{E})$ with the same set of nodes[1]. This two-layer network can be viewed as a cyber-physical system. In current distribution systems, however, the communication links are relatively scarce so that the physically connected buses are not necessarily strongly-connected over the communication network. This scenario is also justified by the fact that the DERs are usually connected into micro-grids, which may be owned by several different entities/controllers. Due to competition or privacy concerns, there is very limited information exchange among the micro-grids. Within each connected area, however, there exists a central controller that can communicate and manage the DERs. Hence, as shown in Fig. 1, we consider a *locally connected communication network* based on a non-overlapping partition of the nodes, where the cyber connectivity within each area follows a star topology. Buses within each of such communication subgraphs constitute a *communication area*, or referred to as *area* in the rest of the paper. The signals, including measurements and decision variables, are exchanged between the controller and the DERs in the area via the local communication network of a star topology. More details on this signal exchanging process are provided in the algorithm design part in Section IV.

Suppose there are $K$ communication areas in $(\mathcal{N}, \mathcal{E}_c)$ and let $\mathcal{K} = \{1, \cdots, K\}$. Let $\mathcal{K}_\kappa$ denote the set of buses within the $\kappa$-th communication area, $\forall \kappa \in \mathcal{K}$. Clearly, $\bigcup_{\kappa \in \mathcal{K}} \mathcal{K}_\kappa = \mathcal{N}_p$ and $\mathcal{K}_{\kappa_1} \bigcap \mathcal{K}_{\kappa_2} = \emptyset, \forall \kappa_1, \kappa_2 \in \mathcal{K}$. Here by little abuse of notation, we use $\mathbf{x}_\kappa \in \mathbb{R}^{2|\mathcal{K}_\kappa|}$ to denote the vector of controllable powers at buses within area-$\kappa$, and similarly for $\mathbf{z}_\kappa \in \mathbb{R}^{m|\mathcal{K}_\kappa|}$ as the controlled vector. Hence, the per-area counterpart of (1) becomes

$$\mathbf{z}_\kappa = \mathbf{H}_{\kappa,\kappa} \mathbf{x}_\kappa + \mathbf{H}_{\kappa,-\kappa} \mathbf{x}_{-\kappa} + \bar{\mathbf{z}}_\kappa \qquad (3)$$

where the subscript $-\kappa$ represents the indices of buses that are not in communication area-$\kappa$. Moreover, $\mathbf{H}_{\kappa,\kappa}$ and $\mathbf{H}_{\kappa,-\kappa}$ are thus submatrices taken from $\mathbf{H}$ with corresponding rows/columns and $\mathbf{x}_{-\kappa}$ denotes the controllable powers of all buses outside area-$\kappa$. Note that the matrix $\mathbf{H}$ is usually dense, especially in voltage regulation problem [10], and so are the submatrices $\mathbf{H}_{\kappa,\kappa}$ and $\mathbf{H}_{\kappa,-\kappa}$.

---
[1]Mathematically speaking, this leads to the fact that $(\mathcal{N}, \mathcal{E}_c)$ is a subgraph of $(\mathcal{N}, \mathcal{E})$.

### III. GAME CHARACTERIZATION UNDER LOCAL COMMUNICATIONS

In this section, we consider a game-theoretic characterization of the DSM problem $\mathcal{P}^{(t)}$ for a given communication network $(\mathcal{N}, \mathcal{E}_c)$. For ease of exposition, we drop the superscript $(t)$ in this section since we first focus on a per-time slot analysis[2].

To solve the problem $\mathcal{P}$ in a distributed fashion, it is natural to decouple the overall objective (2a) into separate communication areas and coordinately solve the centralized problem using distributed/consensus optimization algorithms as in [1], [4]. Nonetheless, under the locally connected $(\mathcal{N}, \mathcal{E}_c)$, no information can be exchanged among communication areas, there is no way for the decision variables and measurement outside area-$\kappa$ to diffuse to the area. This makes it impossible to apply the existing algorithms directly. Although several distributed optimization algorithms have been designed to be robust to link failures that cause instantaneous disconnected communication graph [35]–[37], none has considered a totally disconnected one. It is not clear about the performance of these algorithms under our disconnected architecture of information sharing.

Under our setting, each communication area can only manage its local objective cost, though they are all physically connected and coupled. This characteristic makes the problem fall under the realm of a strategic game [38]. Consider a strategic game $\mathcal{G} = \langle \mathcal{K}, \{\mathcal{X}_\kappa\}_{\kappa \in \mathcal{K}}, \{U_\kappa\}_{\kappa \in \mathcal{K}} \rangle$ with $K$ players, where each player represents one communication area. The *action set* of each player-$\kappa$ is the feasible set of controllable powers $\mathcal{X}_\kappa, \forall \kappa \in \mathcal{K}$. Let $U_\kappa : \mathcal{X} \to \mathbb{R}$ denote the *payoff function* for area-$\kappa$ and $\mathcal{X} := \prod_{\kappa=1}^{K} \mathcal{X}_\kappa$. Similar to (2a), each area aims to minimize its own management cost, and thus $U_\kappa$ takes the following form[3]:

$$U_\kappa(\mathbf{x}) = \frac{\eta}{2} \|\mathbf{z}_\kappa - \boldsymbol{\mu}_\kappa\|_2^2 + \gamma \sum_{j \in \mathcal{K}_\kappa} C_j(\mathbf{x}_j) \qquad (4)$$

$$= \frac{\eta}{2} \|\mathbf{H}_{\kappa,\kappa} \mathbf{x}_\kappa + \mathbf{H}_{\kappa,-\kappa} \mathbf{x}_{-\kappa} + \bar{\boldsymbol{\mu}}_\kappa\|_2^2 + \gamma \sum_{j \in \mathcal{K}_\kappa} C_j(\mathbf{x}_j)$$

---

[2]The notations with time slot index $t$ will be restored later when a dynamic DSM algorithm is proposed in Section V.

[3]Note that here we use the convention that players *minimize* (not maximize) their payoff (cost) functions as in [39].

where $\bar{\boldsymbol{\mu}}_\kappa := \bar{\mathbf{z}}_\kappa - \boldsymbol{\mu}_\kappa$ and clearly, $U := \sum_{\kappa \in \mathcal{K}} U_\kappa$ is the overall cost in (2a).

One significant solution concept to analyze a strategic game is the Nash equilibrium (NE) [38]. An NE point $\mathbf{x}^*$ of the game $\mathcal{G}$ is defined as the solution to the following NE problem (NEP)

$$U_\kappa(\mathbf{x}_\kappa^*, \mathbf{x}_{-\kappa}^*) \leq U_\kappa(\mathbf{x}_\kappa, \mathbf{x}_{-\kappa}^*), \forall \mathbf{x}_\kappa \in \mathcal{X}_\kappa, \forall \kappa \in \mathcal{K} \qquad (5)$$

The following proposition guarantees the existence of such an NE of our game $\mathcal{G}$.

**Proposition 1.** *[Existence of the NE] The set of Nash equilibria of the game $\mathcal{G}$ is nonempty.*

**Proof.** By definition, $U_\kappa(\mathbf{x}_\kappa, \mathbf{x}_{-\kappa})$ is continuously differentiable with respect to (w.r.t.) $\mathbf{x}$ and convex w.r.t. $\mathbf{x}_\kappa$ for any fixed $\mathbf{x}_{-\kappa}$. Moreover, since the action set $\mathcal{X}_\kappa$ is compact and convex, we conclude that the set of NE is nonempty [40]. ∎

Let $\boldsymbol{\Phi}(\mathbf{x}) : \mathcal{X} \to \mathbb{R}^N$ denote the *pseudo-gradient* mapping of the payoff functions $U_\kappa(\mathbf{x}), \forall \kappa \in \mathcal{K}$, which is defined as

$$\begin{aligned}\boldsymbol{\Phi}(\mathbf{x}) &:= [\nabla_{\mathbf{x}_\kappa} U_\kappa(\mathbf{x})]_{\kappa \in \mathcal{K}} \\ &= \eta \tilde{\mathbf{H}}^T (\mathbf{H}\mathbf{x} + \bar{\boldsymbol{\mu}}) + \gamma \nabla \mathbf{C}(\mathbf{x}),\end{aligned} \qquad (6)$$

where $\nabla \mathbf{C}(\mathbf{x}) := [\nabla C_j(\mathbf{x}_j)]_{j \in \mathcal{N}_p}$ is the gradient of the cost function that is decomposable over the powers $\mathbf{x}_j, \forall j \in \mathcal{N}_p$. The matrix $\tilde{\mathbf{H}} := \text{diag}\{\mathbf{H}_{\kappa,\kappa}\} \in \mathbb{R}^{N \times 2N}$ is the block diagonal matrix of $\mathbf{H}$ composed of $\mathbf{H}_{\kappa,\kappa}$, reflecting the partition of the buses in $\mathcal{N}_p$ into communication areas.

**Remark 3.** Note that if all the buses are connected to a central controller, there will be one single area and $\tilde{\mathbf{H}}$ will reduce to $\mathbf{H}$. In this case, the game-theoretic characterization here coincides with that of a network-wide optimization problem as in [8]. If there is no communication link among DERs, $\tilde{\mathbf{H}}$ becomes a diagonal matrix and it recovers the fully local control setting as in [41].

Under Assumption 1, we first obtain the following lemma about the mapping $\boldsymbol{\Phi}$.

**Lemma 1.** *The mapping $\boldsymbol{\Phi}$ is $L_{\boldsymbol{\Phi}}$-Lipschitz continuous, where $L_{\boldsymbol{\Phi}} = \eta \|\tilde{\mathbf{H}}^T \mathbf{H}\|_2 + \gamma L$.*

To obtain stronger results for the uniqueness of the NE, we first consider the case where $\mathcal{X}_j$ are box constraints. Due to the convexity of $U_\kappa(\mathbf{x}_\kappa, \mathbf{x}_{-\kappa})$ w.r.t. $\mathbf{x}_\kappa$, the NEP (5) becomes a box-constrained quadratic programming for area-$\kappa$ given other areas' controllable powers $\mathbf{x}_{-\kappa}$. Assuming the interior point of $\mathcal{X}_\kappa$ is nonempty for each area-$\kappa$, then the solution of each NEP given $\mathbf{x}_{-\kappa}$ can be equivalently characterized by the Karush-Kuhn-Tucker (KKT) conditions [42]. Concatenating the KKT conditions leads to the following equilibrium conditions (EC) that characterize the NE as a whole

$$\begin{cases} \boldsymbol{\Phi}(\mathbf{x}) + \bar{\boldsymbol{\lambda}} - \underline{\boldsymbol{\lambda}} = \mathbf{0} & (7a) \\ \bar{\boldsymbol{\lambda}}^T (\mathbf{x} - \bar{\mathbf{x}}) = 0 & (7b) \\ \underline{\boldsymbol{\lambda}}^T (\mathbf{x} - \underline{\mathbf{x}}) = 0 & (7c) \\ \bar{\boldsymbol{\lambda}} \geq \mathbf{0}, \underline{\boldsymbol{\lambda}} \geq \mathbf{0} & (7d) \\ \underline{\mathbf{x}} \leq \mathbf{x} \leq \bar{\mathbf{x}} & (7e) \end{cases}$$

where $\bar{\boldsymbol{\lambda}}$ and $\underline{\boldsymbol{\lambda}}$ are the multipliers regarding $\bar{\mathbf{x}}_j$ and $\underline{\mathbf{x}}_j$.

If we set $\gamma = 0$, the EC (7) can be viewed as the solution condition of a linear complementarity problem (LCP) as in [43]. By virtue of the theory of LCP, we obtain the following sufficient and necessary condition for the uniqueness of the NE for any choice of $\gamma \geq 0$ and any operating point $\bar{\boldsymbol{\mu}} \in \mathbb{R}^N$.

**Theorem 1.** *[Uniqueness of the NE] For any choice of $\gamma \geq 0$, the NE of the game $\mathcal{G}$ is unique for any operating point captured by $\bar{\boldsymbol{\mu}}$ if and only if the matrix $\tilde{\mathbf{H}}^T\mathbf{H}$ is a P-matrix, i.e., every principal minor of $\tilde{\mathbf{H}}^T\mathbf{H}$ is positive.*

**Proof.** We prove the proposition by extending the proof of Theorem 4.2. in [43]. First note $\mathbf{M} \in \mathbb{R}^{2N \times 2N}$ is a P-matrix if and only if [43]

$$x_i(\mathbf{M}\mathbf{x})_i \leq 0, \forall i = 1, \cdots, 2N \implies \mathbf{x} = \mathbf{0}. \quad (8)$$

**Necessity:** As in [43], one can find a vector $\mathbf{y} \in \mathcal{X}$ such that the solution to the LCP is nonunique if property (8) fails. In our case, we can choose $\mathbf{M} = \tilde{\mathbf{H}}^T\mathbf{H}$ and $\mathbf{y} = (\mathbf{x} - \underline{\mathbf{x}})/\varepsilon$ with sufficiently small $\varepsilon > 0$ such that $\mathbf{x} - \underline{\mathbf{x}} < \bar{\mathbf{x}} - \underline{\mathbf{x}}$ always holds. Then the same counterexample of non-uniqueness applies even for the extreme choice of $\gamma = 0$. Hence $\tilde{\mathbf{H}}^T\mathbf{H}$ has to be a P-matrix for any choice of $\gamma \geq 0$ and $\eta > 0$.

**Sufficiency:** If $\mathbf{x}^1$ and $\mathbf{x}^2$ are both solutions to the EC (7), then there are nine possible combinations of each pair of $x_i^1$ and $x_i^2$'s values: they can be either $\underline{x}_i$, $\bar{x}_i$, or between $(\underline{x}_i, \bar{x}_i)$. Note that the value of $\mathbf{x}$ uniquely determines the signs of $\bar{\boldsymbol{\lambda}}$ and $\underline{\boldsymbol{\lambda}}$ due to the complementarity slackness condition, which further determines the sign of each element of $\boldsymbol{\Phi}(\mathbf{x})$, i.e.,

$$x_i = \underline{x}_i \implies [\boldsymbol{\Phi}(\mathbf{x})]_i \geq 0$$
$$x_i = \bar{x}_i \implies [\boldsymbol{\Phi}(\mathbf{x})]_i \leq 0$$
$$\underline{x}_i < x_i < \bar{x}_i \implies [\boldsymbol{\Phi}(\mathbf{x})]_i = 0.$$

It can be verified that $(\mathbf{x}^1 - \mathbf{x}^2)_i [\boldsymbol{\Phi}(\mathbf{x}^1) - \boldsymbol{\Phi}(\mathbf{x}^2)]_i \leq 0, \forall i$ for all the nine combinations of $x_i^1$ and $x_i^2$. Then by Assumption 1, $(\mathbf{x}^1 - \mathbf{x}^2)_i [\tilde{\mathbf{H}}^T\mathbf{H}(\mathbf{x}^1 - \mathbf{x}^2)]_i \leq 0$ holds due to the convexity of $C_j$ and the positiveness of $\eta$. Therefore the property (8) implies that $\mathbf{x}^1 - \mathbf{x}^2 = \mathbf{0}$, which concludes the uniqueness. ∎

Note that the sufficient and necessary condition in Theorem 1 applies to any choice of $\gamma \geq 0$ and operating point $\bar{\boldsymbol{\mu}}$, which is a relatively strong and strict statement. In fact, for certain $\gamma$ and $\bar{\boldsymbol{\mu}}$ in practice, the NE can be unique even if the matrix $\tilde{\mathbf{H}}^T\mathbf{H}$ is not a P-matrix. One drawback of the condition in Proposition 1 is that it is not easy to verify a P-matrix in general, without evaluating all its principal minors [44]. Therefore, we introduce a more common and easy-to-verify sufficient condition for the uniqueness of the NE for general convex and compact action sets $\mathcal{X}_j, \forall j \in \mathcal{N}_p$ [39].

**Lemma 2.** *The NE of the game $\mathcal{G}$ is unique if the mapping $\boldsymbol{\Phi}$ is strongly monotone.*

The proof of Lemma 2 follows directly from the equivalence between the NE problem and a variational inequality (VI) problem as shown in [39].

## IV. ASYNCHRONOUS DSM ALGORITHM

In this section, we propose a distributed DSM algorithm that can achieve the equilibrium in the *time-invariant* setting as in Section III. For lack of communication among areas, we develop a *projected-gradient*-based algorithm with update relying only on the real-time measurement within each communication area. Moreover, since strict synchronization among areas is impractical under such a disconnected information structure, two asynchronous variants of the algorithm with different assumptions on asynchronism are advocated.

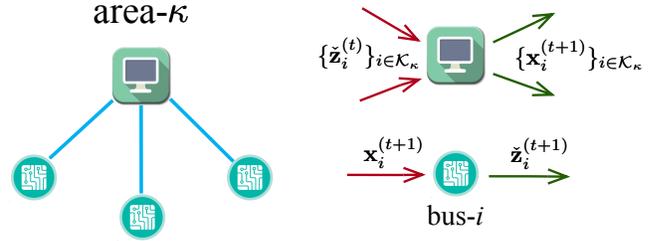

Fig. 2. (left) The locally connected communication network in area-$\kappa$. (right) The signals processed and exchanged between the central controller and the buses in the area.

### A. Algorithm design

First note that the NEP (5) for area-$\kappa$ cannot be solved explicitly at each iteration since other areas' decision $\mathbf{x}_{-\kappa}$ cannot be obtained at area-$\kappa$ via the communication network $(\mathcal{N}, \mathcal{E}_c)$. Thus the best-response-based algorithms [45], [46] are not implementable in our setting. Motivated by the *better-response* play proposed in [47], we develop a projected-gradient-based algorithm for learning the NE with rationality.

For area-$\kappa$, the update at iteration $t$ is based on the instantaneous pseudo-gradient $\boldsymbol{\Phi}_\kappa(\mathbf{x}^{(t)})$, with a given step-size $\epsilon > 0$

$$\mathbf{x}_\kappa^{(t+1)} = \mathbb{P}_\kappa \left[ \mathbf{x}_\kappa^{(t)} - \epsilon \boldsymbol{\Phi}_\kappa(\mathbf{x}^{(t)}) \right]$$
$$= \mathbb{P}_\kappa \left\{ \mathbf{x}_\kappa^{(t)} - \epsilon \left[ \eta \mathbf{H}_{\kappa,\kappa}^T (\mathbf{H}_{\kappa,\kappa}\mathbf{x}_\kappa^{(t)} + \mathbf{H}_{\kappa,-\kappa}\mathbf{x}_{-\kappa}^{(t)} + \bar{\boldsymbol{\mu}}_\kappa) \right. \right.$$
$$\left. \left. + \gamma \nabla \mathbf{C}_\kappa(\mathbf{x}_\kappa^{(t)}) \right] \right\}$$

where the operator $\mathbb{P}_\kappa$ projects any input onto the feasible set of controllable powers $\mathcal{X}_\kappa$. Note that since the matrix $\mathbf{H}$ is usually dense, it seems that the evaluation of the pseudo-gradient $\boldsymbol{\Phi}_\kappa$ still requires other areas' decision $\mathbf{x}_{-\kappa}$. Thanks to the coupling from the physical/power network, the pseudo-gradient can be obtained from the instantaneous measurement of controlled variable $\mathbf{z}$, i.e., $\check{\mathbf{z}}^{(t)}$, following the linear model (1), which leads to an implementable update in a compact form:

$$\mathbf{x}^{(t+1)} = \mathbb{P}\left\{ \mathbf{x}^{(t)} - \epsilon \left[ \eta \tilde{\mathbf{H}}^T (\check{\mathbf{z}}^{(t)} - \boldsymbol{\mu}) + \gamma \nabla \mathbf{C}(\mathbf{x}^{(t)}) \right] \right\}. \quad (9)$$

This update (9) respects the topology of the locally connected communication network. In particular, the central controller of, say area $\kappa$, collects the instantaneous measurement $\{\check{\mathbf{z}}_i^{(t)}\}_{i \in \mathcal{K}_\kappa}$ from the buses within its communication area, in order to update the control decision variables $\mathbf{x}_\kappa^{(t+1)}$. The latter is then sent to each bus via the star-connected communication network and implemented as the control signal. The scheme for signal processing and exchanging in the algorithm is illustrated in Fig. 2.

For the update (9) to converge, we make the following assumption on the mapping $\boldsymbol{\Phi}$ hereafter.

**Assumption 2.** *The pseudo-gradient mapping $\boldsymbol{\Phi}$ is $m$-strongly monotone, i.e., there exists $m > 0$, such that $\forall \mathbf{x}_1, \mathbf{x}_2 \in \mathcal{X}$,*

$$\langle \boldsymbol{\Phi}(\mathbf{x}_1) - \boldsymbol{\Phi}(\mathbf{x}_1), \mathbf{x}_1 - \mathbf{x}_2 \rangle \geq m \cdot \|\mathbf{x}_1 - \mathbf{x}_2\|_2^2.$$

**Remark 4.** We note that the monotonicity of $\boldsymbol{\Phi}$ serves as an essential assumption for most optimization algorithms, or more generally the algorithms for solving variational inequality problems, to converge. For more discussions on this, we refer to [48]. Unlike the projected-gradient update as in [22], the matrix before $\mathbf{x}^{(t)}$ in the pseudo-gradient $\boldsymbol{\Phi}$ is $\tilde{\mathbf{H}}^T \mathbf{H}$, which is generally non-symmetric and may not be positive semi-definite. Even though, since the cost functions $C_{p,j}^{(t)}$ and $C_{q,j}^{(t)}$ are strongly convex by Assumption 1, the strong monotonicity of the mapping $\boldsymbol{\Phi}$ can still be guaranteed with sufficiently large $\gamma > 0$. Additionally, Assumption 2 also ensures the uniqueness of the NE directly from Lemma 2.

For the time-invariant setting, the convergence condition for the update (9) has been investigated in [49], which is stated in the following Proposition.

**Proposition 2.** *Under Assumption 2 and for the time-invariant setting, the update (9) that requires no communication among areas converges to the unique NE point $\mathbf{x}^*$ if the step-size $\epsilon \in (0, 2m/L_{\boldsymbol{\Phi}}^2)$.*

First, the fixed point of the update (9) is the solution to the EC (7), see [50, Prop. 5.1 in Ch. 3]. Then by Lemma 1 and Assumption 2, for the proof it suffices to show that the update (9) is a contraction mapping with small enough $\epsilon$ [49]. Note that due to the non-symmetry of the Jacobian of $\boldsymbol{\Phi}$, the result is weaker than the result for convex optimization problems, which requires $\epsilon \in (0, 2/L_{\boldsymbol{\Phi}})$ as shown in [13]. In fact, only monotonicity (not strong monotonicity) is not sufficient for the convergence when the operator $\boldsymbol{\Phi}$ is non-symmetric [50].

### B. Asynchronous update with bounded delay

Even though the signal exchange can be synchronized by the controller per area, with no inter-area communications, it is challenging to have the updates in (9) synchronized across multiple areas. Thus, we consider two types of asynchronous update where the areas with better computation and sensing capabilities need not wait the slower ones. We first assume the asynchronous update has *bounded delay*, which is a standard setting for many asynchronous algorithms [13], [50], [51]. In particular, the update has the modified form:

$$\mathbf{x}^{(t+1)} = \mathbb{P}\left[\mathbf{x}^{(t)} - \epsilon \mathbf{A}^{(t)} \boldsymbol{\Phi}(\mathbf{x}^{(t)})\right], \tag{10}$$

where $\mathbf{A}^{(t)} \in \mathbb{R}^{2N \times 2N}$ is a block diagonal matrix with $\mathbf{A}_{\kappa,\kappa}^{(t)} \in \mathbb{R}^{2|\mathcal{K}_\kappa| \times 2|\mathcal{K}_\kappa|}$ being the rows and columns of $\mathbf{A}^{(t)}$ indexed by the buses in area-$\kappa$, satisfying

$$\mathbf{A}_{\kappa,\kappa}^{(t)} = \begin{cases} \mathbf{I}_{2|\mathcal{K}_\kappa| \times 2|\mathcal{K}_\kappa|}, & t \in \mathcal{T}_\kappa \\ \mathbf{0}, & t \notin \mathcal{T}_\kappa. \end{cases}$$

where $\mathbf{I}_{a \times b}$ denotes the identity matrix of dimension $a \times b$. The set $\mathcal{T}_\kappa$ collects all the time slots when the area-$\kappa$ conducts the decentralized update. The following standard assumption on $\mathcal{T}_\kappa$ is required for the convergence of the update (10).

**Assumption 3.** *[Boundedness of Update Delay] For every communication area-$\kappa$ and time slot $t$, there exists a positive integer $T$ such that at least one element in the set $\{t, \cdots, t + T - 1\}$ belongs to $\mathcal{T}_\kappa$, i.e., every area must update at least once every $T$ iterations.*

Then the convergence result for the asynchronous update is stated as follows.

**Theorem 2.** *Under Assumption 3 and for the time-invariant setting, the asynchronous update (10) converges to the unique NE point $\mathbf{x}^*$ if the step-size $\epsilon \in (0, 2m/L_{\boldsymbol{\Phi}}^2)$.*

**Proof.** As shown in [50, Prop. 5.1 in Ch. 3], the unique NE point $\mathbf{x}^*$ is the solution to the following fixed point equation

$$\mathbf{x}^* = \mathbb{P}\left[\mathbf{x}^* - \epsilon \boldsymbol{\Phi}(\mathbf{x}^*)\right], \forall \epsilon > 0.$$

Hence if $t \in \mathcal{T}_\kappa$,

$$\begin{aligned} &\|\mathbf{x}_\kappa^{(t+1)} - \mathbf{x}_\kappa^*\|_2^2 \\ &= \left\|\mathbb{P}_\kappa[\mathbf{x}_\kappa^{(t)} - \epsilon \boldsymbol{\Phi}_\kappa(\mathbf{x}^{(t)})] - \mathbb{P}_\kappa[\mathbf{x}_\kappa^* - \epsilon \boldsymbol{\Phi}_\kappa(\mathbf{x}^*)]\right\|_2^2 \\ &\leq \left\|\mathbf{x}_\kappa^{(t)} - \mathbf{x}_\kappa^* - \epsilon[\boldsymbol{\Phi}_\kappa(\mathbf{x}^{(t)}) - \boldsymbol{\Phi}_\kappa(\mathbf{x}^*)]\right\|_2^2 \\ &= \left\|\mathbf{x}_\kappa^{(t)} - \mathbf{x}_\kappa^*\right\|_2^2 - 2\epsilon \langle \mathbf{x}_\kappa^{(t)} - \mathbf{x}_\kappa^*, \boldsymbol{\Phi}_\kappa(\mathbf{x}^{(t)}) - \boldsymbol{\Phi}_\kappa(\mathbf{x}^*) \rangle \\ &\quad \epsilon^2 \left\|\boldsymbol{\Phi}_\kappa(\mathbf{x}^{(t)}) - \boldsymbol{\Phi}_\kappa(\mathbf{x}^*)\right\|_2^2 \\ &\leq (1 - 2\epsilon m + \epsilon^2 L_{\boldsymbol{\Phi}}^2) \left\|\mathbf{x}_\kappa^{(t)} - \mathbf{x}_\kappa^*\right\|_2^2 \end{aligned} \tag{11}$$

where the first inequality comes from the non-expansiveness of the projection operator and the second one follows from Assumption 2 and Lemma 1, that is, the $m$-strong monotonicity and the $L_{\boldsymbol{\Phi}}$-Lipschitz continuity of $\boldsymbol{\Phi}$. Trivially if $t \notin \mathcal{T}_\kappa$, then $\|\mathbf{x}_\kappa^{(t+1)} - \mathbf{x}_\kappa^*\|_2^2 \leq \|\mathbf{x}_\kappa^{(t)} - \mathbf{x}_\kappa^*\|_2^2$. Combining this with (11) we have that $\forall t$,

$$\|\mathbf{x}^{(t+1)} - \mathbf{x}^*\|_2^2 \leq \|\mathbf{x}^{(t)} - \mathbf{x}^*\|_2^2 - \rho(\epsilon) \|\mathbf{A}^{(t)}(\mathbf{x}^{(t)} - \mathbf{x}^*)\|_2^2 \tag{12}$$

where

$$\rho(\epsilon) := 2\epsilon m - \epsilon^2 L_{\boldsymbol{\Phi}}^2. \tag{13}$$

By summing up the inequalities (12) for all $t$, we have

$$\sum_{t=0}^{\infty} \|\mathbf{A}^{(t)}(\mathbf{x}^{(t)} - \mathbf{x}^*)\|_2^2 \leq \rho^{-1}(\epsilon) \|\mathbf{x}^{(0)} - \mathbf{x}^*\|_2^2$$

provided $\rho(\epsilon) > 0$, which equivalently requires $\epsilon \in (0, 2m/L_{\boldsymbol{\Phi}}^2)$. Hence the sequence $\{\mathbf{x}_\kappa^{(t)}\}$ converges with the limit $\lim_{t \to \infty} \mathbf{A}_{\kappa,\kappa}^{(t)}(\mathbf{x}_\kappa^{(t)} - \mathbf{x}_\kappa^*) = \mathbf{0}$ for every $\kappa \in \mathcal{K}$, and so does its sub-sequence $\{\mathbf{x}_\kappa^{(t_\kappa)}\} \subseteq \{\mathbf{x}_\kappa^{(t)}\}$ with $t_\kappa \in \mathcal{T}_\kappa$. This means that $\lim_{t_\kappa \to \infty} \mathbf{x}_\kappa^{(t_\kappa)} = \mathbf{x}_\kappa^*$ since $\mathbf{A}_{\kappa,\kappa}^{(t_\kappa)} = \mathbf{I}_{2|\mathcal{K}_\kappa| \times 2|\mathcal{K}_\kappa|}$. On the other hand, from (12), the sequence $\{\|\mathbf{x}^{(t)} - \mathbf{x}^*\|_2^2\}$ is non-increasing and lower bounded by 0. Thus it must converge to a limit point, which has to be 0 in this case since its sub-sequences $\{\mathbf{x}_\kappa^{(t_\kappa)}\}, \forall \kappa \in \mathcal{K}$ satisfy $\lim_{t_\kappa \to \infty} \|\mathbf{x}_\kappa^{(t_\kappa)} - \mathbf{x}_\kappa^*\|_2^2 = 0$. This leads to the conclusion that $\lim_{t \to \infty} \mathbf{x}^{(t)} = \mathbf{x}^*$. ∎

Note that unlike the proof in [13] for algorithms solving an optimization problem, no global objective function is defined explicitly in our game-theoretic setting. Instead, we find $\|\mathbf{x}^{(t+1)} - \mathbf{x}^*\|_2^2$ as a valid Lyapunov function to show the convergence. Similar to the result in [13], the step-size $\epsilon$ for learning the Nash equilibrium of the game is not compromised by the asynchronous update with bounded delay. This is attributed to the fact that the measurement from the physical layer always provides the latest pseudo-gradient information.

### C. Asynchronous random update

Consider another type of asynchronous update with randomness, where whether each communication area-$\kappa$ updates or not at time slot $t$ follows a Bernoulli distribution. This random update is possibly due to the random link failure within each area, which has been used to model noisy links as in sensor networks [52] and communication networks [53]. To be specific, we have the following assumption for the random update.

**Assumption 4.** *[Bernoulli Random Update] At each time slot $t$, the communication area-$\kappa$ chooses to update its powers $\mathbf{x}^{(t)}$ following an independent and identically distributed (i.i.d.) Bernoulli distribution with mean $\bar{b}_\kappa > 0$, i.e., the overall update for all areas follows*

$$\mathbf{x}^{(t+1)} = \mathbb{P}\left[\mathbf{x}^{(t)} - \epsilon \mathbf{B}^{(t)} \mathbf{\Phi}(\mathbf{x}^{(t)})\right], \quad (14)$$

*where $\mathbf{B}^{(t)} \in \mathbb{R}^{2N \times 2N}$ is a random block diagonal matrix with $\mathbf{B}^{(t)}_{\kappa,\kappa}$ on its diagonal, satisfying*

$$\mathbf{B}^{(t)}_{\kappa,\kappa} = \begin{cases} \mathbf{I}_{2|\mathcal{K}_\kappa| \times 2|\mathcal{K}_\kappa|}, & \text{with prob } \bar{b}_\kappa \\ \mathbf{0}, & \text{with prob } 1 - \bar{b}_\kappa. \end{cases} \quad (15)$$

*Also, the random $\mathbf{B}^{(t)}$ is independent of the update $\mathbf{x}^{(t)}$.*

We thus obtain the following convergence statement under this asynchronous random update.

**Theorem 3.** *Under Assumption 4, the asynchronous update (14) converges linearly to the unique NE point $\mathbf{x}^*$ in the mean-square sense (m.s.s.) if the the step-size $\epsilon \in (0, 2m/L_\mathbf{\Phi}^2)$, with the convergence rate $1 - \rho(\epsilon) \cdot \min_{\kappa \in \mathcal{K}}\{\bar{b}_\kappa\}$, where $\rho(\epsilon)$ is defined as in (13).*

**Proof.** First notice the similarity between the update rules (14) and (10), we have the counterpart of (12) by taking expectation over both sides,

$$\mathbb{E}\|\mathbf{x}^{(t+1)} - \mathbf{x}^*\|_2^2$$
$$\leq \mathbb{E}\|\mathbf{x}^{(t)} - \mathbf{x}^*\|_2^2 - \rho(\epsilon)\mathbb{E}\|\mathbf{B}^{(t)}(\mathbf{x}^{(t)} - \mathbf{x}^*)\|_2^2$$
$$= \mathbb{E}\|\mathbf{x}^{(t)} - \mathbf{x}^*\|_2^2 - \rho(\epsilon)\mathbb{E}\|\mathbf{x}^{(t)} - \mathbf{x}^*\|_{\bar{\mathbf{B}}}^2$$
$$\leq \left(1 - \rho(\epsilon)\min_{\kappa \in \mathcal{K}}\{\bar{b}_\kappa\}\right) \cdot \mathbb{E}\|\mathbf{x}^{(t)} - \mathbf{x}^*\|_2^2, \quad (16)$$

where

$$\bar{\mathbf{B}} := \mathbb{E}\left[\mathbf{B}^{(t)T}\mathbf{B}^{(t)}\right] \quad (17)$$

is a block diagonal matrix with the diagonal matrix $\bar{\mathbf{B}}_{\kappa,\kappa} = \bar{b}_\kappa \cdot \mathbf{I}_{2|\mathcal{K}_\kappa| \times 2|\mathcal{K}_\kappa|}$. The second equality of (16) comes from

$$\mathbb{E}\|\mathbf{B}^{(t)}(\mathbf{x}^{(t)} - \mathbf{x}^*)\|_2^2$$
$$= \mathbb{E}[(\mathbf{x}^{(t)} - \mathbf{x}^*)^T \mathbf{B}^{(t)T} \mathbf{B}^{(t)} (\mathbf{x}^{(t)} - \mathbf{x}^*)]$$
$$= \mathbb{E}\{\text{Tr}[(\mathbf{x}^{(t)} - \mathbf{x}^*)(\mathbf{x}^{(t)} - \mathbf{x}^*)^T \mathbf{B}^{(t)T} \mathbf{B}^{(t)}]\}$$
$$= \text{Tr}\{\mathbb{E}[(\mathbf{x}^{(t)} - \mathbf{x}^*)(\mathbf{x}^{(t)} - \mathbf{x}^*)^T] \cdot \mathbb{E}[\mathbf{B}^{(t)T} \mathbf{B}^{(t)}]\}$$
$$= \mathbb{E}[(\mathbf{x}^{(t)} - \mathbf{x}^*)^T \bar{\mathbf{B}} (\mathbf{x}^{(t)} - \mathbf{x}^*)] \quad (18)$$

since $\mathbf{x}^{(t)}$ and $\mathbf{B}^{(t)}$ are independent. Therefore, (16) leads to the linear convergence rate of the sequence $\{\mathbf{x}^{(t)}\}$ provided that $\bar{b}_\kappa$ is bounded below from 0 and $\rho(\epsilon) > 0$, or equivalently, $\epsilon \in (0, 2m/L_\mathbf{\Phi}^2)$. ∎

Note that the requirement for a convergent step-size $\epsilon$ remains the same as in Proposition 2 and Theorem 2. Additionally, linear rate of convergence is guaranteed in the m.s.s., in contrast to the update with bounded delay that ensures only asymptotic convergence. As we will show shortly, the linear rate with constant step-size facilitates the convergence of the algorithm in the dynamic setting.

## V. DYNAMIC DSM ALGORITHM

To achieve the final goal of the dynamic DSM, we extend the earlier analysis to *time-varying* setting which is attributed to the inevitable volatility of load and renewable generation in distribution systems. We also investigate the impact of measurement mismatches of $\mathbf{z}$ on the tracking performance of the algorithm.

Consider a time-varying characterization of the game $\mathcal{G}^{(t)}$

$$\mathcal{G}^{(t)} = \left\langle \mathcal{K}, \left\{\mathcal{X}_\kappa^{(t)}\right\}_{\kappa \in \mathcal{K}}, \left\{U_\kappa^{(t)}\right\}_{\kappa \in \mathcal{K}} \right\rangle$$

with time-varying action sets and payoff functions. The set $\mathcal{X}_\kappa^{(t)}$ reflects the time-varying capability of DERs' controllable powers affected by the external environment such as weather conditions. The payoff $U_\kappa^{(t)}$ has the same form as (4) while subject to time-varying operating point $\bar{\boldsymbol{\mu}}_\kappa^{(t)}$ and cost function $C_j^{(t)}(\mathbf{x}_j)$. Under Assumption 2, the Nash Equilibrium, denoted as $\mathbf{x}^{*,(t)}$, is unique at each time $t$. We first state the following assumption on the time-varying NE points to make them trackable.

**Assumption 5.** *[Boundedness of NE Point Drift] The successive difference of the transient NE point of the game $\mathcal{G}^{(t)}$ is bounded, i.e., there exists a positive constant $B_1$ such that $\forall t$,*

$$\mathbb{E}\|\mathbf{x}^{*,(t+1)} - \mathbf{x}^{*,(t)}\|_2^2 \leq B_1.$$

This assumption would hold under bounded difference of parameters that define consecutive games, including $\bar{\boldsymbol{\mu}}_\kappa^{(t)}$, $C_j^{(t)}(\mathbf{x}_j)$, and $\mathcal{X}_\kappa^{(t)}$. Note that no assumption on the evolution dynamics of the settings is required, as opposed to the first-order autoregressive (AR(1)) process assumed in [13].

For practical implementation of the asynchronous update (14), the real-time measurements $\check{\mathbf{z}}^{(t)}$ are used as advocated in (9). Nonetheless, the performance of the implementation may suffer from the measurement mismatches between $\check{\mathbf{z}}^{(t)}$ and

$\mathbf{z}^{(t)}$. Note that the mismatch accounts for both i) the sensing errors of the DERs and ii) the modeling errors due to the linearization of a possibly nonlinear model as (1). Here we make a standard assumption that the measurement mismatch is also bounded.

**Assumption 6.** *[Boundedness of Measurement Mismatches] The real-time measurement mismatch of $\mathbf{z}$ is bounded, i.e., there exists a positive constant $B_2$ such that $\forall t = 0, 1, \cdots$,*

$$\mathbb{E}\|\check{\mathbf{z}}^{(t)} - \mathbf{z}^{(t)}\|_2^2 \leq B_2,$$

*where $\mathbf{z}^{(t)}$ is determined by the powers $\mathbf{x}^{(t)}$ following power flow equation* (2b).

Now we are ready to present the tracking properties of the asynchronous random update (14) for the dynamic setting.

**Theorem 4.** *Define*

$$\xi(\epsilon) := (1+c_2)(1+c_1)\big[1 - \rho(\epsilon) \cdot \min_{\kappa \in \mathcal{K}}\{\bar{b}_\kappa\}\big]$$
$$\Delta(\epsilon) := \left(1 + \frac{1}{c_2}\right)\left(1 + \frac{1}{c_1}\right)\epsilon^2 \eta^2 \|\bar{\mathbf{B}}\|_2 \|\tilde{\mathbf{H}}\|_2^2 B_2 + B_1,$$

*$\forall c_1, c_2 > 0$. Under Assumptions 4, 5 and 6, for any small $c_1, c_2$ such that $\xi(\epsilon) \in (0,1)$, if $\epsilon \in (0, 2m/L_\Phi^2)$, we have*

$$\mathbb{E}\|\mathbf{x}^{(t)} - \mathbf{x}^{*,(t)}\|_2^2 \leq [\xi(\epsilon)]^t \mathbb{E}\|\mathbf{x}^{(0)} - \mathbf{x}^{*,(0)}\|_2^2 + \frac{1-[\xi(\epsilon)]^t}{1-\xi(\epsilon)}\Delta(\epsilon).$$

**Proof.** Define first the time-varying mappings

$$\mathbf{\Phi}^{(t)}(\mathbf{x}) := \eta\tilde{\mathbf{H}}^T(\mathbf{z} - \boldsymbol{\mu}) + \gamma \nabla \mathbf{C}^{(t)}(\mathbf{x})$$
$$\check{\mathbf{\Phi}}^{(t)}(\mathbf{x}) := \eta\tilde{\mathbf{H}}^T(\check{\mathbf{z}} - \boldsymbol{\mu}) + \gamma \nabla \mathbf{C}^{(t)}(\mathbf{x}),$$

where $\mathbf{z}$ is uniquely determined by $\mathbf{x}$ and $\check{\mathbf{z}}$ is the measurement of $\mathbf{z}$. Hence the asynchronous random update using real-time measurements becomes

$$\mathbf{x}^{(t+1)} = \mathbb{P}^{(t)}\left[\mathbf{x}^{(t)} - \epsilon \mathbf{B}^{(t)} \check{\mathbf{\Phi}}^{(t)}(\mathbf{x}^{(t)})\right]. \quad (19)$$

Denote the update with exact $\mathbf{z}$ as $\tilde{\mathbf{x}}^{(t+1)}$, then $\tilde{\mathbf{x}}^{(t+1)} := \mathbb{P}^{(t)}[\mathbf{x}^{(t)} - \epsilon \mathbf{B}^{(t)} \mathbf{\Phi}^{(t)}(\mathbf{x}^{(t)})]$.

Consider the difference between $\mathbf{x}^{(t+1)}$ and $\mathbf{x}^{*,(t)}$, using the fact that $\|\mathbf{a}+\mathbf{b}\|_2^2 \leq (1+c)\|\mathbf{a}\|_2^2 + (1+1/c)\|\mathbf{b}\|_2^2$ for any $c > 0$, we obtain that $\forall c_1 > 0$

$$\mathbb{E}\|\mathbf{x}^{(t+1)} - \mathbf{x}^{*,(t)}\|_2^2$$
$$\leq (1+c_1)\mathbb{E}\|\tilde{\mathbf{x}}^{(t+1)} - \mathbf{x}^{*,(t)}\|_2^2 + \left(1+\frac{1}{c_1}\right)\mathbb{E}\|\mathbf{x}^{(t+1)} - \tilde{\mathbf{x}}^{(t+1)}\|_2^2$$
$$\leq (1+c_1)\xi'(\epsilon)\mathbb{E}\|\mathbf{x}^{(t)} - \mathbf{x}^{*,(t)}\|_2^2 + \left(1+\frac{1}{c_1}\right)\mathbb{E}\|\mathbf{x}^{(t+1)} - \tilde{\mathbf{x}}^{(t+1)}\|_2^2, \quad (20)$$

where $\xi'(\epsilon) := 1 - \rho(\epsilon) \cdot \min_{\kappa \in \mathcal{K}}\{\bar{b}_\kappa\}$ and the second inequality uses (16). The second term on the right hand side of (20) can be further bounded

$$\mathbb{E}\|\mathbf{x}^{(t+1)} - \tilde{\mathbf{x}}^{(t+1)}\|_2^2 \leq \epsilon^2 \mathbb{E}\|\mathbf{B}^{(t)}[\check{\mathbf{\Phi}}^{(t)}(\mathbf{x}^{(t)}) - \mathbf{\Phi}^{(t)}(\mathbf{x}^{(t)})]\|_2^2$$
$$= \epsilon^2 \mathbb{E}\|\check{\mathbf{\Phi}}^{(t)}(\mathbf{x}^{(t)}) - \mathbf{\Phi}^{(t)}(\mathbf{x}^{(t)})\|_{\bar{\mathbf{B}}}^2$$
$$\leq \epsilon^2 \|\bar{\mathbf{B}}\|_2 \mathbb{E}\|\check{\mathbf{\Phi}}^{(t)}(\mathbf{x}^{(t)}) - \mathbf{\Phi}^{(t)}(\mathbf{x}^{(t)})\|_2^2$$
$$\leq \epsilon^2 \eta^2 \|\bar{\mathbf{B}}\|_2 \|\tilde{\mathbf{H}}\|_2^2 B_2, \quad (21)$$

where the first inequality is due to the non-expansiveness of projection $\mathbb{P}$ and the equality is due to the independence between $\mathbf{B}^{(t)}$ and $\mathbf{x}^{(t)}$ with $\bar{\mathbf{B}}$ defined as in (17). The last inequality is based on Assumption 6.

By combining (20) and (21), we have $\forall c_2 > 0$

$$\mathbb{E}\|\mathbf{x}^{(t+1)} - \mathbf{x}^{*,(t+1)}\|_2^2$$
$$\leq (1+c_2) \cdot \mathbb{E}\|\mathbf{x}^{(t+1)} - \mathbf{x}^{*,(t)}\|_2^2 + \left(1+\frac{1}{c_2}\right)\mathbb{E}\|\mathbf{x}^{*,(t+1)} - \mathbf{x}^{*,(t)}\|_2^2$$
$$\leq (1+c_2)(1+c_1)\xi'(\epsilon) \cdot \mathbb{E}\|\mathbf{x}^{(t)} - \mathbf{x}^{*,(t)}\|_2^2 +$$
$$\left(1+\frac{1}{c_2}\right)\left(1+\frac{1}{c_1}\right)\epsilon^2\eta^2\|\bar{\mathbf{B}}\|_2\|\tilde{\mathbf{H}}\|_2^2 B_2 + B_1$$
$$\leq [\xi(\epsilon)]^{t+1} \cdot \mathbb{E}\|\mathbf{x}^{(0)} - \mathbf{x}^{*,(0)}\|_2^2 + \frac{1-[\xi(\epsilon)]^{t+1}}{1-\xi(\epsilon)} \cdot \Delta(\epsilon). \quad (22)$$

Note that since $\xi'(\epsilon) < 1$, $c_1$ and $c_2$ can be arbitrarily small such that $\xi(\epsilon) < 1$, which completes the proof. ■

Theorem 4 establishes the performance of the real-time update (19) on tracking the time-varying NE point. It shows that with the same choice of step-size as for the *time-invariant* and *synchronous* setting, the tracking error vanishes at an exponential rate. We also immediately obtain the following corollary about the steady-state tracking error.

**Corollary 1.** *[Steady-state Tracking Error Bound] Under the conditions of Theorem 4, the steady-state tracking error for the dynamic setting is bounded as*

$$\limsup_{t \to \infty} \mathbb{E}\|\mathbf{x}^{(t)} - \mathbf{x}^{*,(t)}\|_2^2 \leq \frac{\Delta(\epsilon)}{1-\xi(\epsilon)}. \quad (23)$$

Corollary 1 shows that the size of the neighborhood of the NE where the dynamic DSM update converges to is bounded eventually. Note that the impact of the step-size on the bound for steady-state tracking error is more involved than the results in [13]. In fact, i) due to the requirement for strong monotonicity of $\mathbf{\Phi}$, the geometric rate $\xi(\epsilon)$ is a quadratic (not monotone) function of the step-size $\epsilon$; ii) due to the measurement mismatch of $\check{\mathbf{z}}^{(t)}$, the constant error $\Delta$ is also a function of $\epsilon$. We will show several examples to illustrate this in Section VI.

**Remark 5.** Compared to the results in [13], we generally make improvement in three aspects: i) we provide performance analysis under two dynamic scenarios, asynchronous update and time-varying operating conditions, simultaneously; ii) no assumption on the dynamic model of the operating point, e.g., the AR(1) process, as assumed in [13], is required for concluding the tracking performance; iii) the impact of measurement and model errors on the tracking performance is discussed. Moreover, in contrast to the *online OPF* work [27], we consider the limited communication structure that is more realistic in current distribution systems.

## VI. NUMERICAL TESTS

In this section, we investigate the performance of the dynamic DSM algorithm by applying it to a standard DSM task, the voltage regulation problem, where the controlled quantity $\mathbf{z}$ represents the voltage magnitude across the network.

## A. Simulation setup

We first perform numerical tests on the IEEE 13-bus feeder case[4] for distribution systems [54] as illustrated in Fig. 3. Each bus is assumed to have a DER with controllable real and reactive powers. On top of this physical layer, we consider four cases of communication networks, which partition the buses into areas as shown in Table I. Test Case 1 reduces to the case with a strongly connected communication network and one area, while Test Case 4 corresponds to the fully local case with no communication among DERs. The impedance of each line segment is set as $(0.233+j0.366)\Omega$, under the base of 4.16 kV and 100 kVA. We choose $\eta = 1$, $\gamma = 2$, and the desired voltage $\mu = 1$. For the dynamic setting, we set the cost function $C_j^{(t)}(\mathbf{x}_j) = c_{p,j}\|p_j - p_{j,max}^{(t)}\|_2^2/2 + c_{q,j}\|q_j\|_2^2/2$, with $c_{p,j} = c_{q,j} = 10^{-3}$ and $p_{j,max}^{(t)}$ being the maximum power available at the DER as in [27]. The vector of $p_{j,max}^{(t)}, \forall j$, denoted as $\mathbf{p}_{max}^{(t)}$, is modeled as an AR(1) process

$$\mathbf{p}_{max}^{(t+1)} = \mathbf{F}_\mathbf{p} \mathbf{p}_{max}^{(t)} + \boldsymbol{\delta}_\mathbf{p}^{(t)} + \mathbf{c}_\mathbf{p},$$

where $\mathbf{c}_\mathbf{p}$ is a constant vector, $\mathbf{F}_\mathbf{p} := \alpha_\mathbf{p} \mathbf{I}$ is a time-invariant transition matrix with $\alpha_\mathbf{p} \in (0,1)$ referred to as the *forgetting factor*, and $\boldsymbol{\delta}_\mathbf{p}^{(t)}$ represents a zero-mean white noise process with covariance matrix $\boldsymbol{\Sigma}_\mathbf{p}$. In addition, the feasible set of controllable powers at each bus is $\mathcal{X}_j^{(t)} = [-1.0 + \delta_{\bar{\mathbf{x}}_j}^{(t)}, 1.0 + \delta_{\underline{\mathbf{x}}_j}^{(t)}]$ p.u., where both $\delta_{\bar{\mathbf{x}}_j}^{(t)}$ and $\delta_{\underline{\mathbf{x}}_j}^{(t)}$ are modeled as zero-mean white noise processes with a common covariance matrix $\boldsymbol{\Sigma}_{\mathbf{x}_j}$. The covariance $\boldsymbol{\Sigma}_{\mathbf{x}_j}$ is determined by the environment and weather conditions, for example, the solar irradiance for DERs with PV inverters. As for the time-varying operating point $\bar{\mathbf{z}}^{(t)}$, the AR(1) process is also advocated as in [13]. In particular,

$$\bar{\mathbf{z}}^{(t+1)} = \mathbf{F}_{\bar{\mathbf{z}}} \bar{\mathbf{z}}^{(t)} + \boldsymbol{\delta}_{\bar{\mathbf{z}}}^{(t)} + \mathbf{c}_{\bar{\mathbf{z}}},$$

where $\mathbf{c}_{\bar{\mathbf{z}}}$ is a constant vector, $\mathbf{F}_{\bar{\mathbf{z}}} := \alpha_{\bar{\mathbf{z}}} \mathbf{I}$ is the transition matrix with $\alpha_{\bar{\mathbf{z}}} \in (0,1)$ and $\boldsymbol{\delta}_{\bar{\mathbf{z}}}^{(t)}$ is a zero-mean white noise process with covariance matrix $\boldsymbol{\Sigma}_{\bar{\mathbf{z}}}$. The measurement $\check{\mathbf{z}}^{(t)}$ is modeled as $\check{\mathbf{z}}^{(t+1)} = \mathbf{z}^{(t)} + \boldsymbol{\delta}_{\check{\mathbf{z}}}^{(t)}$ where $\boldsymbol{\delta}_{\check{\mathbf{z}}}^{(t)}$ is a zero-mean white noise process with covariance $\boldsymbol{\Sigma}_{\check{\mathbf{z}}}$.

With this setting, the pseudo-gradient mapping $\boldsymbol{\Phi}$ becomes a linear mapping, of which the strong monotonicity is determined by the matrix $\tilde{\mathbf{H}}^T \mathbf{H}$, i.e., the topology of communication networks, and the coefficients $c_{p,j}$ and $c_{q,j}$. As demonstrated in Table I, the mapping $\boldsymbol{\Phi}$ satisfies Assumption 2 in all the four cases of communication networks, which leads to the uniqueness of the NE according to Lemma 2. Moreover, the modeling of time-varying quantities enables Assumption 5 to be valid since the stochastic processes all have bounded covariance; see more detailed discussions in [13]. Assumption 6 is also satisfied with the bounded covariance of the measurement mismatch $\boldsymbol{\Sigma}_{\check{\mathbf{z}}}$.

In addition, the proposed algorithms are also tested on a more realistic system model, i.e., the IEEE 123-bus case [54], with communication areas partitioned as in Fig. 4. Moreover, the evolution of the operating point $\bar{\mathbf{z}}^{(t)}$ is generated from the real data of time-series load profile from the European

[4]Note that we renumber the buses for both 13-bus and 123-bus cases for ease of reference.

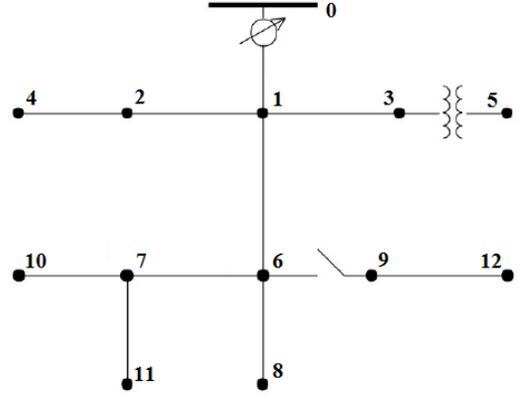

Fig. 3. One-line diagram of the IEEE 13-bus test feeder case.

TABLE I
SUMMARY OF THE FOUR CASES OF COMMUNICATION NETWORKS WITH VARIOUS PARTITIONS OF THE BUSES.

| Test Case | # of Comp. | Partitioned Sets of Buses |
|---|---|---|
| 1 | 1 | $\{1,\cdots,12\}$ |
| 2 | 3 | $\{1,\cdots,5\},\{6,8,9,12\},\{7,10,11\}$ |
| 3 | 5 | $\{1,2,4\},\{3,5\},\{6,8\},\{7,10,11\},\{9,12\}$ |
| 4 | 12 | $\{1\},\{2\},\cdots,\{12\}$ |

| Test Case | $m$ of $\phi$ | Cost | LOE |
|---|---|---|---|
| 1 | 0.002 | 0.0106 | 0 |
| 2 | 0.002 | 0.0121 | 14.15% |
| 3 | 0.002 | 0.0133 | 25.47% |
| 4 | 0.002 | 0.0168 | 58.49% |

Low Voltage Test feeder system with 1-minute resolution [54]. Fig. 5 illustrates the evolution of $\bar{\mathbf{z}}_j^{(t)}$ at three buses during a day, which are determined by the real data of load following the power flow equation (1). The evolution of the maximum available power $\mathbf{p}_{max}^{(t)}$, the feasible set of controllable powers $\mathcal{X}_j^{(t)}$ at DERs are modeled similar to that in the 13-bus case, i.e., $\mathbf{p}_{max}^{(t)}$ follows an AR(1) process, $\mathcal{X}_j^{(t)} = [-1.0 + \delta_{\bar{\mathbf{x}}_j}^{(t)}, 1.0 + \delta_{\underline{\mathbf{x}}_j}^{(t)}]$, with $\delta_{\bar{\mathbf{x}}_j}^{(t)}$ and $\delta_{\underline{\mathbf{x}}_j}^{(t)}$ as white noises.

## B. Value of communication links

A static setting with time-invariant game-theoretic characterization is first considered to evaluate the performance limits of DSM with a locally connected communication network. Specifically, the conditions (7) are solved for the NE of the game directly. Following the techniques in [55], the nonlinear KKT conditions can be transformed to a mixed integer linear programming (MILP) and attacked readily by standard solvers as Gurobi [56].

To quantify the value of communication links, we define the following quantity, named as loss of efficiency (LOE)

$$\text{LOE} = \frac{\sum_{\kappa \in \mathcal{K}} U_\kappa(\mathbf{x}^*) - \sum_{\kappa \in \mathcal{K}} U_\kappa(\mathbf{x}^o)}{\sum_{\kappa \in \mathcal{K}} U_\kappa(\mathbf{x}^o)},$$

where $\mathbf{x}^*$ represents the powers injection at the NE and $\mathbf{x}^o$ represents that at the optimum when the communication network is strongly connected. LOE captures the relative difference between the operational cost at the NE and that at the optimum, which reflects the value of communication links in DSM problems. As shown in Table I, the fewer communication areas in the system, that is, the more communication links are constructed, the higher overall benefit the buses have. Also note that the change of communication topologies, i.e., the change of $\tilde{\mathbf{H}}$, does not change the strong convexity of the mapping $\mathbf{\Phi}$ nor the uniqueness of the NE.

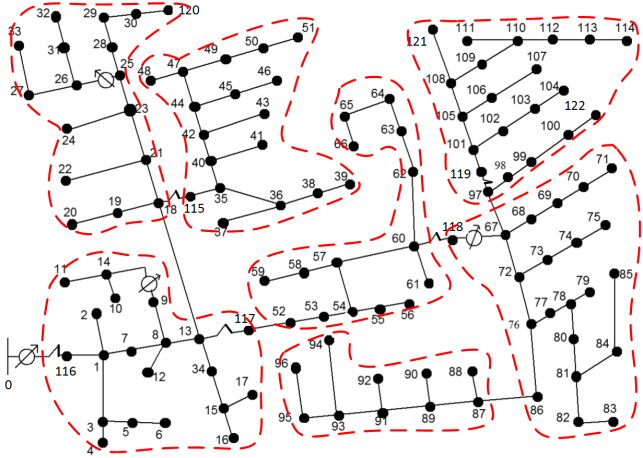

Fig. 4. One-line diagram of the IEEE 123-bus test feeder case with partitioned communication areas. The red dashed lines circle out the 7 communication areas.

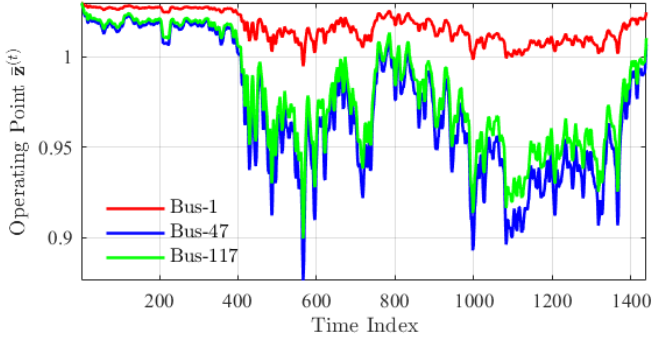

Fig. 5. The evolution of time-varying operating point $\bar{v}_j^{(t)}$ during a day at bus-1, bus-47, bus-117, which are determined by the real load data from the European Low Voltage Test feeder system [54].

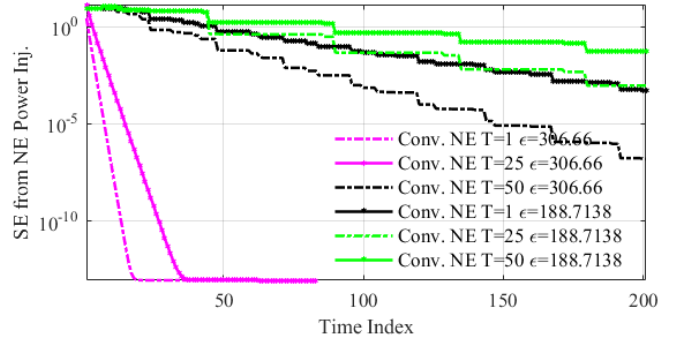

Fig. 6. SE of the iterative powers injection from the NE powers injection using asynchronous updates with bounded delay for Test Case 3.

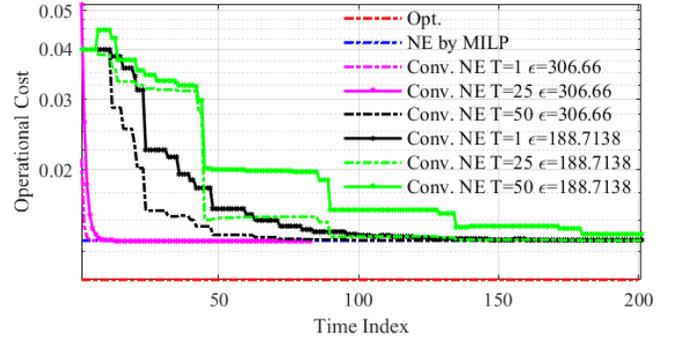

Fig. 7. Iterative operational costs for the asynchronous update with bounded delay for Test Case 3.

By virtue of the asynchronous algorithms developed in Section IV, the NE can be achieved with no communication among areas. Fig. 6 shows the convergence of the squared-error (SE) $\|\mathbf{x}^{(t)} - \mathbf{x}^*\|_2^2$ using the asynchronous updates with bounded delay for Test Case 3. The convergence of operational cost to that at the NE is illustrated in Fig. 7. The step-size $\epsilon$ is set as $\epsilon < 2m/L_{\mathbf{\Phi}}^2 = 409.0353$ following Theorem 2 and enables the updates to converge to the NE as desired. Both figures corroborate the convergence of the algorithm to the Nash equilibrium with small enough step-sizes. Smaller delay bounds $T$ and larger step-sizes $\epsilon$ lead to faster convergence of the asynchronous update. Thanks to the real-time voltage measurement, the choice of step-size is not affected by the delay bound $T$ and is more efficient than the choice using the classical convergence conditions for asynchronous updates as in [50]. Moreover, the LOE, i.e., the gap between the of operational cost at the optimum and that at the NE, is also verified in Fig. 7.

The asynchronous random updates with activation rate following Bernoulli distribution at each area are also evaluated. The parameters $\bar{b}_\kappa, \forall \kappa$ are randomly chosen from uniform distribution over $[\bar{b}-0.05, \bar{b}+0.05]$, where the average value $\bar{b}$ is selected as 0.1, 0.2, and 0.5, respectively. The same choice of step-sizes are employed for successful convergence in the m.s.s. Fig. 8 and Fig. 9 plot the averaged mean-square-error (MSE) from the powers injection at the NE, i.e., $\mathbb{E}\|\mathbf{x}^{(t)} - \mathbf{x}^*\|_2^2$, over 30 random realizations of the update paths over time. It is illustrated that without compromising the choice of step-size, the powers injection converges to the equilibrium injection in both Test Cases at a linear rate. This corroborates the theoretical results we have derived in Section IV. Moreover, larger step-sizes and a larger update probability $\bar{b}_k$ correspond to faster convergence rate.

### C. Asynchronous update under dynamic setting

To verify the convergence results under the dynamic setting, we generate the time-varying quantities as introduced in VI-A. We set $\alpha_{\mathbf{p}} = \alpha_{\bar{\mathbf{z}}} = 0.1$, the evolution-related noise covariance $\mathbf{\Sigma}_{\mathbf{p}} = \mathbf{\Sigma}_{\mathbf{x}_j} = \mathbf{\Sigma}_{\bar{\mathbf{z}}} = \sigma_{evol}^2 \mathbf{I}$ with $\sigma_{evol}^2 = 1 \times 10^{-4}$, while the measurement noise covariance $\mathbf{\Sigma}_{\tilde{\mathbf{z}}} = \sigma_{meas}^2 \mathbf{I}$ with $\sigma_{meas}^2 = 1 \times 10^{-4}$ as the benchmark setting. The mean value of the probability for Bernoulli random update, i.e., $\bar{b}$, is set as 0.2. All the simulation results as follows are averaged over 30 random realizations of update paths over time.

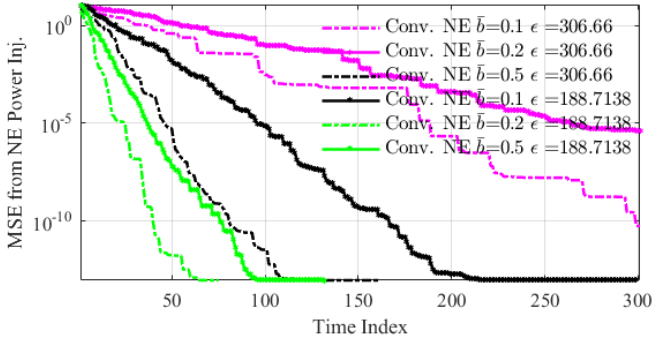

Fig. 8. MSE of the iterative powers injection from the NE powers injection using asynchronous random updates for Test Case 2.

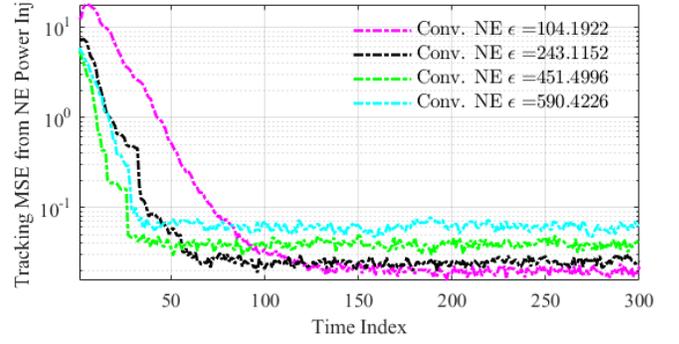

Fig. 10. Tracking error of the powers injection using asynchronous random updates with various step-sizes $\epsilon$ for Test Case 3 under dynamic setting.

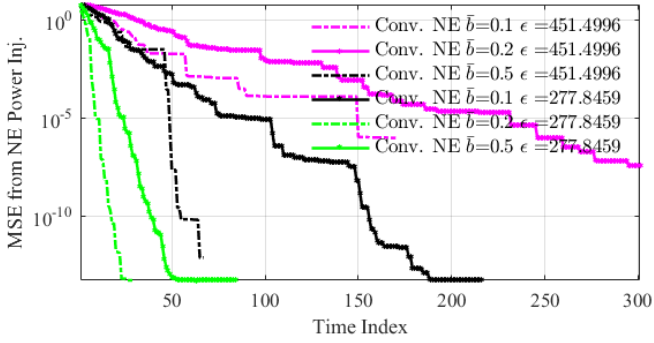

Fig. 9. MSE of the iterative powers injection from the NE powers injection using asynchronous random updates for Test Case 3.

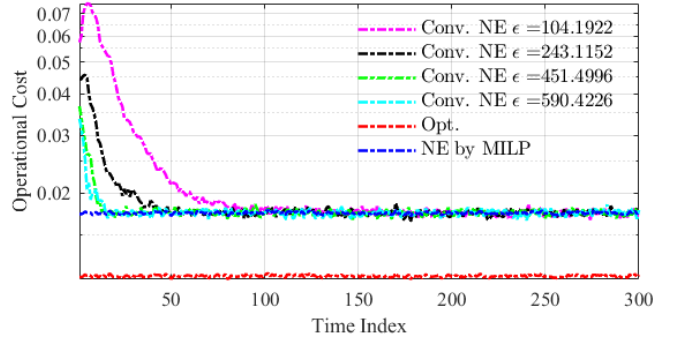

Fig. 11. Iterative operational costs for the asynchronous random update with various step-sizes $\epsilon$ for Test Case 3 under dynamic setting.

*1) The step-size:* We first investigate the impact of step-size $\epsilon$ on the convergence speed and the steady-state tracking error. The step-sizes are chosen as the coefficient $[0.3, 0.7, 1.3, 1.7]/2$ times the maximum convergent step-size $2m/L_{\bar{\Phi}}^2$, which equals 694.6148 for Test Case 4. As shown in Fig. 10, the steady-state tracking error is indeed non-zero but upper-bounded as proved in Theorem 4. It is also shown in Fig. 10 that for step-sizes smaller than $m/L_{\bar{\Phi}}^2$, the larger $\epsilon$ is, the faster the asynchronous update converges. For step-sizes greater than $m/L_{\bar{\Phi}}^2$, however, a larger step-size does not benefit the convergence speed. This is due to the fact that $\eta(\epsilon)$ is quadratic with respect to $\epsilon$ with minimum taken at $m/L_{\bar{\Phi}}^2$. Moreover, it seems that the steady-state tracking error increases monotonically with the increase of step-sizes in this Test Case. Therefore, with faster transient evolutions, $\epsilon$ should be chosen around $m/L_{\bar{\Phi}}^2$; while for smaller steady-state tracking errors, a relatively smaller step-size is preferred.

Interestingly, although the MSE of $\mathbf{x}^{(t)}$ does not vanish to zero, the updates still converge to the NE in terms of operational cost as shown in Fig. 11. This is mainly due to the insensitivity of the cost w.r.t. the controllable powers $\mathbf{x}^{(t)}$.

*2) The transient drift of NE:* We then investigate how the drift of the NE point impacts the convergence and steady-state tracking error. According to Lemma 1 in [23], the consecutive difference of AR(1) sequence is determined by $2\sigma^2/(1+\alpha)$, where $\sigma^2 \mathbf{I}$ is the covariance matrix of the noise process and $\alpha$ is the forgetting factor. Note that since $\delta_{\bar{\mathbf{x}}_j}^{(t)}$ and $\delta_{\underline{\mathbf{x}}_j}^{(t)}$ are both white noise processes, they are not considered as a factor that influences the drift of the NE in the m.s.s.

Therefore, we first fix the variance of the AR(1) processes as $\sigma^2/(1-\alpha^2) = 1 \times 10^{-4}$, and vary the forgetting factor of both $\bar{\mathbf{z}}^{(t)}$ and $\mathbf{p}_{max}^{(t)}$ from the benchmark setting. Accordingly, the bound of the drift of the NE $B_1$ should decrease with $\alpha_{\bar{\mathbf{z}}}$ and $\alpha_{\mathbf{p}}$ approaching 1. As shown in Fig. 12, the greater forgetting factor leads to smaller steady-state tracking error along with smaller bound $B_1$.

We then fixed the $\alpha_{\bar{\mathbf{z}}}$ and $\alpha_{\mathbf{p}}$ as 0.1 and vary the evolution variance of the AR models to affect the bound $B_1$. As shown in Fig. 13, a smaller evolution variance will reduce the tracking error at steady-state, as it reduces the term $B_1$ in the tracking error bound in Theorem 4.

*3) The measurement mismatch:* Additionally, we also investigate the effect of measurement mismatch on the convergence results we have derived. As illustrated in Fig. 14, a higher level of measurement noise will lift the steady-state error bound as stated in Theorem 4, as it increases the measurement mismatch bound $B_2$.

To sum up, the numerical test for the four Test Cases verify the analytical results that i) with a locally connected communication network, the loss of efficiency will occur; ii) the convergence speed of the proposed algorithms depend non-monotonically on the choice of the step-size $\epsilon$; iii) the steady-state tracking error performance indeed depends on the proper choice of $\epsilon$ and also affected by the uncertainty of the evolution of the setting and the measurement mismatches. This is consistent with the upper bound derived in Theorem 4.

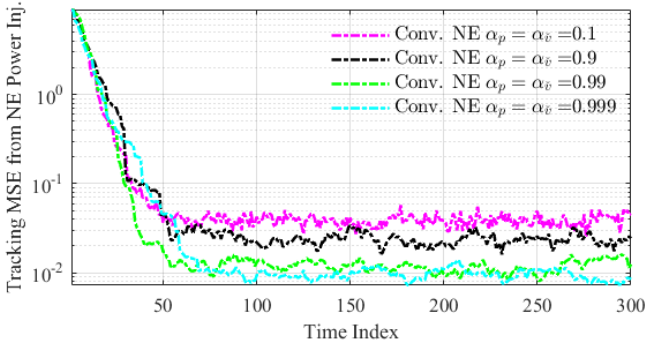

Fig. 12. Tracking error of the powers injection using asynchronous random updates with various forgetting factors $\alpha_{\mathbf{p}}$ and $\alpha_{\bar{\mathbf{z}}}$ for Test Case 3 under dynamic setting.

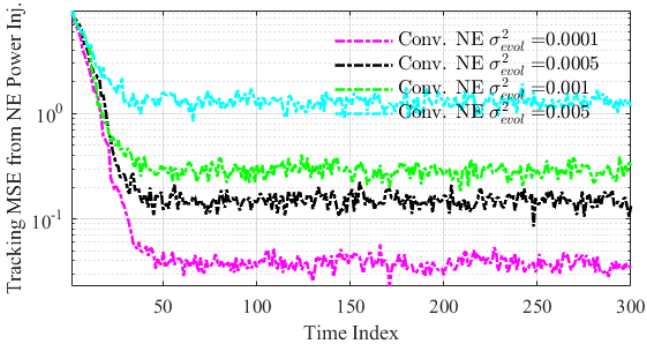

Fig. 13. Tracking error of the powers injection using asynchronous random updates with various evolution covariances $\sigma^2_{evol}$ for Test Case 3 under dynamic setting.

### D. Larger system with real dynamic data

To verify that the convergence results hold regardless of the random process of the operating point $\bar{\mathbf{z}}^{(t)}$, we test our algorithm using the real time-varying voltage profile on a larger system. As shown in Fig. 15, there always exists a gap between the operational cost at the optimum with strongly-connected (*Opt.* in the figure) and that at the NE with locally connected communication network (*NE by MILP* in the figure). Due to the abrupt changes of operating point of the real system, the proposed dynamic algorithm can not track the time-varying NE point as closely as it does for the test data under the AR(1) model. In fact, the steady-state performance can hardly be observed under this non-stationary stochastic process. We demonstrate the tracking performance in Fig. 16. To handle the time-varying setting, the other two algorithms solve the game $G^{(t)}$ in an offline fashion every 15 and 30 minutes, respectively. It has been observed that our dynamic algorithm achieves lower average tracking error (ATE) in contrast to the offline algorithms, with even lighter computation (one step of gradient-projection per minute). This has substantiated the advantage of our online algorithm.

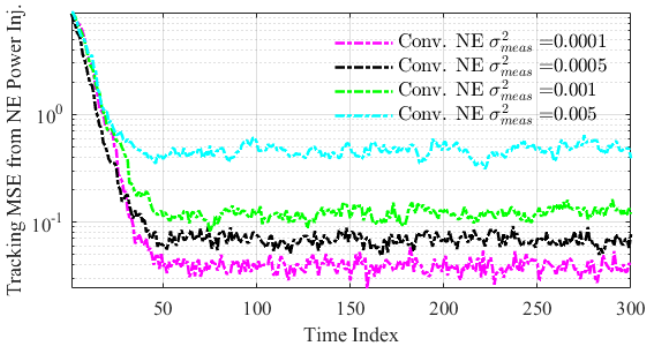

Fig. 14. Tracking error of the powers injection using asynchronous random updates with various measurement mismatch covariances $\sigma^2_{meas}$ for Test Case 3 under dynamic setting.

## VII. CONCLUDING REMARKS

In this paper, we have addressed the challenges in modeling and designing algorithms for dynamic distribution system management using DERs under a locally connected communication network. We have proposed a game-theoretic characterization for this scenario where the DERs are partitioned into several communication areas with only intra-area information exchange allowed. We have then developed a projected-gradient-based distributed algorithm that depends on real-time measurements of the controlled quantity to achieve the Nash equilibrium. We have shown both analytically and numerically that the algorithm can accommodate the dynamic setting, which is potentially induced by the asynchronous updates of DERs, the time-varying operating conditions, and the measurement mismatches concurrently.

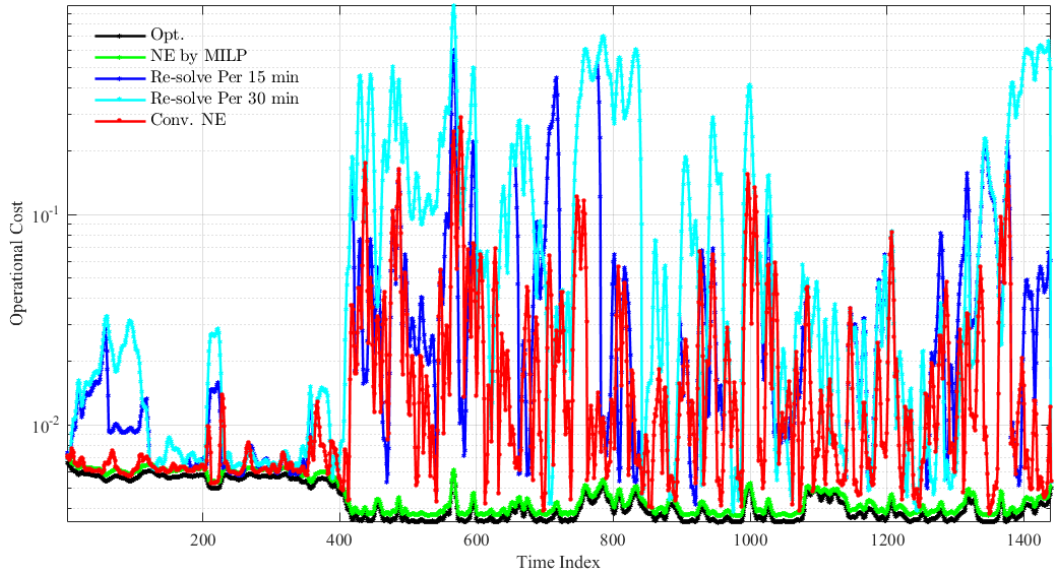

Fig. 15. Tracking error of the powers injection using asynchronous random updates for IEEE 123-bus case under a dynamic setting using real data.

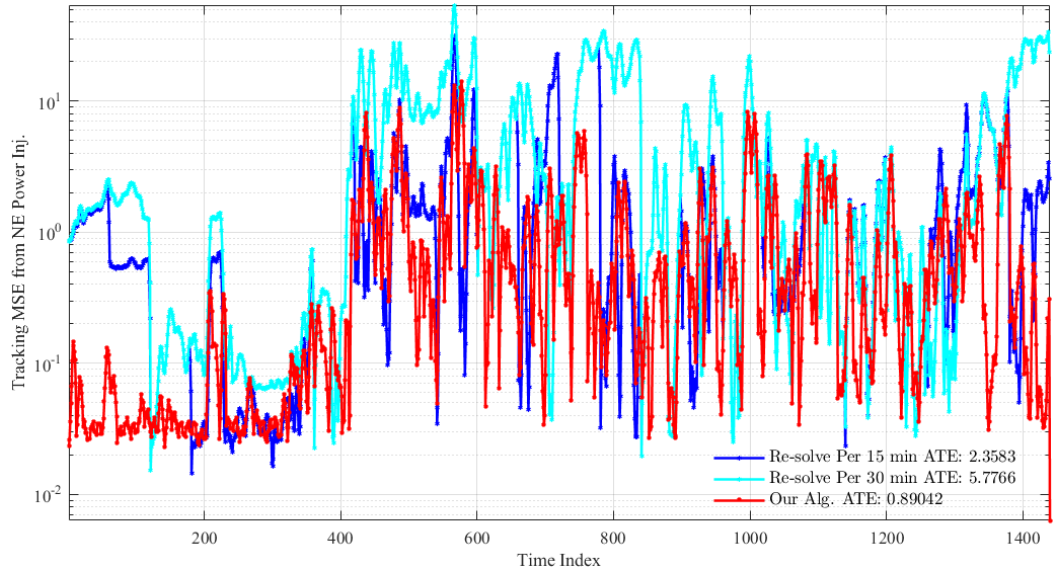

Fig. 16. Iterative operational costs for the asynchronous random update for IEEE 123-bus case under a dynamic setting using real data.